\documentclass[prc,twocolumn]{revtex4}

\usepackage{graphicx}
\usepackage{dcolumn}
\usepackage{bm}
\usepackage{amsmath}

\begin{document}

\title{Elastic Magnetic Electron Scattering from Deformed Nuclei}
\author{P.~Sarriguren}
\email{p.sarriguren@csic.es}
\author{D.~Merino}
\affiliation{
Instituto de Estructura de la Materia, IEM-CSIC, Serrano
123, E-28006 Madrid, Spain}

\author{O.~Moreno}
\author{E.~Moya~de~Guerra}
\affiliation{Departamento de Estructura de la Materia, F\'\i sica T\'ermica y
Electr\'onica, and IPARCOS, Facultad de Ciencias F\'\i sicas, Universidad 
Complutense de Madrid, Madrid E-28040, Spain}

\author{D.~N.~Kadrev}
\author{A.~N.~Antonov}
\author{M.~K.~Gaidarov}
\affiliation{
Institute for Nuclear Research and Nuclear Energy,
Bulgarian Academy of Sciences, Sofia 1784, Bulgaria}

%\date{\today}

\begin{abstract}
Magnetic form factors corresponding to elastic electron scattering from odd-$A$ nuclei 
are presented. The calculations are carried out in plane-wave Born approximation. The 
one-body properties are obtained in a deformed self-consistent mean-field calculation 
based on a Skyrme HF+BCS method. Collective effects are also included in the cranking 
approximation. Results on several stable nuclei are compared with the available 
experimental information. It is shown that a deformed formalism improves the agreement 
with experiment in deformed nuclei, while reproducing equally well spherical nuclei 
by taking properly the spherical limit of the deformed model and the effect of 
nucleon-nucleon correlations. Thus, the capability of the model to describe magnetic 
form factors is demonstrated. This opens the door to explore also unstable nuclei of 
particular interest that could be measured in future experiments on electron-radioactive 
beam colliders. 

\end{abstract}

\maketitle

\section{Introduction}

Electron scattering is a unique tool for studying the electromagnetic properties of 
nuclei, getting insight into the nuclear charge and current distributions. There are 
several reasons for using electrons as probes. First, the electron interacts with the 
nucleus with the electromagnetic force, which is the best known interaction, accurately 
described by quantum electrodynamics. The coupling constant of the interaction is also 
sufficiently weak to not significantly disturb the nuclear structure under study. In 
addition, the weakness of the interaction allows one to work in first order perturbation 
within the one-photon exchange approximation. Secondly, in contrast to the case of real 
photons, one can vary the energy transfer and the momentum transfer independently, thus 
mapping out the Fourier transform of the densities. Extensive work on electron 
scattering has been performed in the past in the elastic, inelastic, and quasielastic 
regimes, providing us with the most accurate measurements on charge radii, transition 
probabilities, as well as  momentum distributions and spectroscopic factors 
\cite{hofstadter,forest,donnwal,vri87,frullani,udias}. 

Electron scattering is not only sensitive to the nuclear charge distribution. Electrons 
also scatter from the nuclear electromagnetic current distributions and the experimental 
observation of this process provides information on the convection and magnetization 
currents within the nucleus. In particular, elastic magnetic electron scattering 
provides fine details of the nuclear ground-state convection and magnetization current 
distributions \cite{donnelly,elvira1,elvira2}.

Although magnetic scattering shares the advantages of the electromagnetic probes, there 
are significant differences with respect to charge scattering. Because the angular 
momenta of the nucleons pair off within the core, the magnetic properties in odd-$A$ 
nuclei are determined to a large extent by the unpaired nucleon. Therefore, magnetic 
scattering mainly provides information on the single-particle properties of the valence 
nuclear wave function, whereas the collective aspects that dominate charge scattering 
show up only in particular cases. In addition, the intrinsic magnetic moments of protons 
and neutrons are quite similar in magnitude and thus, magnetic electron scattering will 
provide information on both, contrary to the case of charge scattering which is mostly 
sensitive to protons. 

Electron scattering experiments have been limited so far to stable isotopes, but the 
advantages of electron probes mentioned above can be exploited to gain information on 
unstable nuclei as well. At present, there is a large activity devoted to investigate 
with such probes the properties of stable \cite{dalinac} and unstable isotopes at 
radioactive nuclear beam facilities \cite{sudasimon}. The technical challenges to 
perform such experiments on unstable nuclei, are being considered in new facilities 
like ELISe (FAIR-GSI) \cite{elise} and SCRIT (RIKEN) \cite{scrit}. The conceptual 
design and the scientific challenges of the electron-ion collider ELISe (Electron-Ion 
Scattering in a Storage Ring) can be found in Ref. \cite{anton_nima}. In SCRIT 
(Self-Confining Radioactive Ion Target), a circulating beam of electrons scatters off 
ions stored in a trap \cite{scrit_npn,ohnishi}, and the first elastic electron 
scattering experiment on $^{132}$Xe has already been performed \cite{scrit_xe}.

New interesting and challenging aspects like nuclear halos and neutron skins can be 
addressed with these new electron scattering experiments from exotic nuclei. More 
generally, the evolution of the charge distributions in isotopic chains can give 
valuable information on the extent to which these phenomena may trigger sizable 
changes in the charge radius. This can be a test of the different theoretical models 
used for predicting charge distributions. Examples of these studies can be found in 
Refs. \cite{garrido,antonov_ff,kara,dong,roca_08,roca_13,liu1,liu2,radhi_18,liang_18}.

Similarly to the case of charge scattering, magnetic scattering has been also studied 
in stable nuclei from different theoretical frameworks \cite{donnelly}, including shell 
model \cite{radhi_03,radhi_07,jassim,samma}, relativistic mean field 
\cite{wang14,wang15,liu16,guo}, and deformed mean field models
\cite{elvira1,elvira2,diep,kova,kova1,graca,ps}.

Since the magnetic form factors are mainly determined by the last occupied orbital of
the nucleon, valuable information on the characteristics of these single-particle
properties have been extracted on stable nuclei. These experiments could become in the 
future an alternative tool to study the valence structure of odd-$A$ exotic nuclei, in 
particular, the wave function of the loosely bound halo nucleons that may be only 
slightly perturbed  by the electromagnetic probe. Electron scattering would be an 
additional tool to complement the information obtained by other means. This is for 
example the case of magnetic dipole (M1) excitations in nuclei, where it is well 
known \cite{heyde,sarri_m1} that complementary information is obtained when using 
different electromagnetic $(\gamma , \gamma '),\, (e,e')$ or hadronic $(p,p')$ probes. 

Deformation can be an important issue when dealing with the structural evolution in 
isotopic chains, including both stable and unstable isotopes, undergoing a shape 
transition. A model describing properly the isotopic evolution of these properties is 
desirable.

The aim of this work is to analyze the dependence on deformation of the magnetic form 
factors of odd-$A$ nuclei and to show the capability of our deformed formalism to 
describe both spherical and deformed nuclei. To fulfill this purpose, we compare our 
results with the available experimental information on stable nuclei. 

The paper is organized as follows. In the next section we present the theoretical
formalism to calculate the magnetic form factors in deformed nuclei, as well as its 
spherical limit. Section III contains the results obtained in the case of spherical 
and deformed nuclei. Section IV contains the conclusions of the work.

\section{Theoretical Formalism}

\subsection{Cross Section and Form Factors}

The formalism of electron scattering from deformed nuclei that we follow in this work
was introduced and discussed earlier \cite{elvira1,diep,kova}. In particular, the
work in \cite{kova,kova1} demonstrated the need for deformation to account for the
data in $^{181}$Ta. Actually, magnetic scattering on $^{181}$Ta was a case of success 
where first came theory  and then experiment \cite{rad_exp} confirmed the theoretical
predictions in \cite{kova}. The method has been already applied to different cases 
\cite{graca,ps,berdi,nk}, where the various sensitivities of the results to different 
approximations concerning nuclear structure and reaction mechanism were studied.
In particular, it was shown that the magnetic form factors of deformed nuclei may 
differ considerably from those of spherical nuclei. 
The sensitivity tests can be further explored, but the focus in this paper is mainly
to investigate the dependence of the form factors on deformation.

Here, we briefly summarize this formalism. Following the notation of Ref. \cite{elvira1}, 
the general cross section for electron scattering of ultrarelativistic electrons for
transitions from the nuclear ground state ($I_i$) to final states ($I_f$) is given in 
the plane-wave Born approximation (PWBA) by

\begin{equation}
\left. \frac{d\sigma}{d\Omega} \right| _{I_i\rightarrow I_f} = 4\pi \sigma_M f_{\rm rec}^{-1}
\left[ V_L |F_L|^2 + V_T |F_T|^2 \right] \, ,
\label{eq1}
\end{equation}
in terms of the Mott cross section

\begin{equation}
\sigma_M = \left[\frac{\alpha \cos(\theta/2)}{2\epsilon_i \sin^2 (\theta/2)}\right]^2 \, ,
\end{equation}
and the recoil factor,

\begin{equation}
f_{\rm rec} = 1+\frac{2\epsilon_i \sin^2(\theta/2)}{M_{\rm Target}}\, .
\end{equation}
The dependence on the electron kinematics is given by the longitudinal and transverse
Rosenbluth factors,

\begin{equation}
V_L= (Q^2/q^2)^2\, , \quad V_T = \tan^2(\theta /2)-(Q^2/2q^2) \, ,
\label{eq4}
\end{equation}
where the kinematical variables are defined so that an incident electron with 
four-momentum $k_{i\mu}=(\epsilon_i,{\bf k_i})$ is scattered through an angle $\theta$ 
to four-momentum $k_{f\mu}=(\epsilon_f,{\bf k_f})$ by exchanging a virtual photon with 
four-momentum $Q=(\omega,{\bf q})$. The cross section can be separated into longitudinal 
(L) or Coulomb, and transverse (T) parts, weighted with different kinematical factors.

Because all the charged nucleons contribute equally to the Coulomb (charge) form 
factors $F_L$, they scale like $Z^2$ in the cross section. On the other hand, transverse 
form factors $F_T$ are basically single-particle observables that depend mostly on the 
properties of the unpaired nucleon in the outermost shell and do not carry this factor. 
As a consequence, $F_L$ dominates at most angles and special kinematic conditions are
needed to maximize magnetic $F_T$ contributions. This is why backward scattering
 ($\theta = 180^{\circ} $) is commonly used to measure $F_T$ (see Eqs. (\ref{eq1}) - 
(\ref{eq4})). If magnetic form factors were to be measured in unstable nuclei, this 
difficulty of backward scattering will add to those inherent to the scattering from 
radioactive beams. In electron scattering experiments from unstable nuclei, the number 
of targets available is an important issue due to the limitations in the number of 
nuclei that can be produced and to their short half-lives. The luminosity ${\cal L}$, 
defined as the ratio of the event rate per time to the cross-section, is a key parameter 
to asses the feasibility of the experiment. Luminosities over $10^{27}$ cm$^{-2}$ s$^{-1}$ 
have been already achieved at the SCRIT facility in the $^{132}$Xe experiment 
\cite{scrit_xe} and similar luminosities are expected for unstable nuclei in the next 
future \cite{ohnishi} that will allow to measure them. The luminosities expected in 
ELISe are also within similar ranges \cite{anton_nima}.

The dependence on the nuclear structure is contained in the $q$-dependent longitudinal
and transverse form factors, which are written in terms of Coulomb (C), transverse
electric (E), and transverse magnetic (M) multipoles,

\begin{equation}
|F_L|^2 = \sum _{\lambda \ge 0} |F^{C\lambda}|^2\, , \quad
|F_T|^2 = \sum _{\lambda \ge 1} \left[ |F^{M\lambda}|^2 + |F^{E\lambda}|^2 \right] ,
\end{equation}
which are defined as the reduced matrix elements of the multipole operators  
$\hat{T}^{\sigma\lambda}$ between initial and final nuclear states

\begin{equation}
|F^{\sigma\lambda}|^2 = \frac{ \left| \langle I_f || \hat{T}^{\sigma\lambda}(q)||I_i
\rangle \right|^2} {2I_i+1} \, .
\label{f_rme} 
\end{equation}

For elastic scattering, parity and time reversal invariance imply that only the even 
Coulomb and odd transverse magnetic multipoles contribute. Then, at $\theta=180^{\circ}$
only odd magnetic multipoles will survive in PWBA,

\begin{equation}
| F_T(q) |^2 = \sum_{\lambda={\rm odd}} | F^{M\lambda} |^2 \, .
\label{ftfm}
\end{equation}

The magnetic multipole operators are defined as 

\begin{equation}
\hat{T}^{M\lambda}_{\mu}(q) = i^{\lambda}  \int d{\bf r}\, j_{\lambda}(qr) 
{\bf Y}^{\mu}_{\lambda\lambda}(\Omega_r) \cdot \hat{\bf{J}}  ({\bf r}) \, ,
\label{tensor}
\end{equation}
where $\hat{\bf{J}} ({\bf r})$ is the current density operator. The currents 
$\hat{\bf{J}}$ in the transverse form factors contain both convection and magnetization 
components that arise from the motion and from the intrinsic magnetic moments of the 
nucleons, respectively.

When we calculate the total form factors, we include center of mass (c.m.) and finite 
nucleon size corrections. For the c.m. correction we use the usual factor obtained in 
the harmonic-oscillator approximation,

\begin{equation}
f_{\rm c.m.}=\exp{\left[ b^2 q^2/(4A) \right] }\, ,
\end{equation}
with the oscillator length $b=A^{1/6}$ fm. 

In the magnetization currents, we use bare nucleon magnetic moments,
$\mu_{s}^p=2.793\ \mu_N$, $\mu_{s}^n=-1.913\ \mu_N$, corrected with dipole form 
factors \cite{sim80},

\begin{equation}
G^M_{\tau}(q^2)=\mu_s^{\tau} [1+q^2/(18.23 \ {\rm fm}^{-2})]^{-2}\, .
\end{equation}

In the convection currents, we use bare orbital $g$-factors, $g_{\ell}^p=1$ and 
$g_{\ell}^n=0$ scaled by $q$-dependent form factors. The proton electric form factor is 
given by a sum of monopoles parametrized in Ref. \cite{sim80}, 

\begin{equation}
G^E_{p}(q^2)=\sum_{n=1}^4 \frac{a_n}{1+q^2/m_n^2}\, .
\end{equation}
The neutron electric form factor is given by the  difference of two 
Gaussians \cite{chan76},

\begin{equation}
G^E_{n}(q^2)=\exp (-q^2r_{+}^2/4) - \exp (-q^2r_{-}^2/4)\, ,
\end{equation}
with $r_\mp ^2=0.507413\pm  0.038664 \ {\rm fm}^2$ .

The effects of Coulomb distortion can be treated in a quantitative way in the 
distorted-wave Born approximation (DWBA) with a phase-shift calculation \cite{yen54}. 
Nevertheless, 
for the analysis and interpretation of experimental data on magnetic scattering, 
neglecting the Coulomb distortion offers clear advantages and PWBA is commonly used. 
In PWBA the connection between data and the underlying physical quantities is more 
transparent, and calculations are much easier. Therefore, it is convenient to convert 
first the experimental magnetic cross sections into plane-wave form factors that can be 
used in PWBA interpretation. The most important effect of Coulomb distortion can be 
accounted for by using an effective momentum transfer. This procedure was done in the 
data used in this paper (taken from Ref. \cite{donnelly}) and therefore, the 
experimental form factors were converted into plane-wave form factors that can be 
directly compared with PWBA calculations.

It is worth noting that, in DWBA, backward electron scattering may receive also
longitudinal contributions that could mix with the transverse ones. Then, a careful 
subtraction of longitudinal contributions should, in principle, be done before
interpreting the backward scattering as purely transverse. These effects were studied 
in Ref. \cite{nishimura} concluding that on heavy deformed nuclei the distortion effects 
enhance somewhat the cross section at low $q$, while PWBA is reliable at larger 
$q$-values beyond 1 fm$^{-1}$. For the medium-mass nuclei and magnetic multipoles 
considered in this work, these effects are not expected to be important.

\subsection{Form Factors in a Deformed Formalism}

The ground state of axially symmetric deformed nuclei is characterized by angular 
momentum $I$, projection along the symmetry axis $k$, and parity $\pi$. Initial and 
final states in Eq. (\ref {f_rme}) are the same for elastic scattering ($Ik^{\pi}$).

The magnetic  $F^{M\lambda}$ multipole form factors can be written in terms of intrinsic 
form factors ${\cal F}^{M \lambda}$ weighted by angular momentum dependent coefficients.
To lowest order in an expansion in powers of the total angular momentum, the intrinsic 
multipoles depend only on the intrinsic structure of the ground-state 
band \cite{elvira1,elvira2}. 
The transition multipoles in Eq. (\ref{ftfm}) for the elastic case $I_f = I_i = k$ can 
be written as

\begin{eqnarray}
\left. F^{M\lambda} \right| _{\rm def}  
&=&\langle kk \lambda 0\ |kk\rangle {\cal F}^{M\lambda}_k 
+ \langle k\ -k\ \lambda \ 2k|kk\rangle {\cal F}^{M\lambda}_{2k} \nonumber \\
&&+\frac{\lambda(\lambda+1)}{\sqrt{2}} 
\langle kk\lambda 0 |kk\rangle {\cal F}^{M\lambda}_R \, .
\label{f_rot}
\end{eqnarray}
${\cal F}^{M \lambda}_R$ are the transverse multipoles of the collective rotational 
current (rotational multipoles) that depend on the nuclear rotational model used to 
describe the band \cite{elvira1,elvira2}. 
The single-particle multipoles ${\cal F}^{M\lambda}_k$ 
and  ${\cal F}^{M \lambda}_{2k}$ depend only on the single-particle intrinsic wave 
function of the odd nucleon if the even-even core is time-reversal invariant, as we 
assume in this work. They are different from zero only for $k\ne 0$ bands and are 
given by,

\begin{eqnarray}
 {\cal F}^{M\lambda}_{k} & = &  \langle \phi_k | \hat{T} ^{M\lambda}_0 |
\phi_k \rangle \, , \label{fmk}\\
{\cal F}^{M\lambda}_{2k} & = &  \langle \phi_k | \hat{T} ^{M\lambda}_{2k} |
\phi_{\bar k} \rangle +\delta_{k,1/2} \frac{a}{\sqrt{2}} {\cal F}^{M\lambda}_{R} \, . 
\label{fm2k}
\end{eqnarray}
$\hat{T} ^{M \lambda}_{\mu}$ is the $\mu$ component of the $M \lambda$ tensor operator,
see Eq. (\ref{tensor}) and Ref. \cite{forest}.  $\phi_k$ and $\phi_{\bar k}$ are the 
wave functions of the odd nucleon and its time reverse, respectively, and 
$a=<\phi_k|j_+|\phi_{\bar k}>$ is the decoupling parameter for $k=1/2$ bands.

As we can see from the above expressions, the magnetic form factors in odd-$A$ nuclei
receive two types of contributions, single-particle and collective. While the former
depends strongly on the single-particle state occupied by the unpaired nucleon, the 
latter depends on the rotational properties of the even-even core. The interplay 
between these two types of contributions and their relative intensities were discussed 
in \cite{ps}. It was found that the single-particle form factors are dominant at most 
$q$-values. They depend on the quantum number $k$ of the band, on the neutron or proton 
character of the odd nucleon, as well as on the mean field used to generate the 
single-particle states. The coupling of the unpaired nucleon to the deformed core and 
the currents connected with the collective rotation play also a role.

In the HF+BCS method for axially symmetric deformed nuclei, the wave function for the 
$i$-state is written in terms of the spin components $\phi_i^+$ and $\phi_i^-$
as \cite{vautherin},
\begin{eqnarray}
\phi_i({\bf R},\sigma) &=&  \phi_i^+ (r,z)
\exp(\mathrm{i}\Lambda^-\varphi) \chi_+(\sigma) \nonumber \\
&& + \phi_i^- (r,z) \exp(\mathrm{i}\Lambda^+\varphi) \chi_-(\sigma) .
\end{eqnarray}
The variables $r$, $z$ and $\varphi$ are the cylindrical coordinates of the 
radius-vector ${\bf R}$. $\chi_\pm(\sigma)$ are the spin wave functions and 
$\Lambda^\pm = \Omega_i \pm 1/2 \ge 0$, where $\Omega_i$ is the projection along the 
symmetry axis of the total angular momentum, and it characterizes the single-particle 
Hartree-Fock solutions for axially symmetric deformed nuclei, together with parity 
$\pi_i$. 

The wave functions $\phi_i$ are expanded into eigenfunctions 
$\psi_\alpha ({\bf R},\sigma)$ of an axially deformed harmonic oscillator potential,
\begin{equation}
\phi_i({\bf R},\sigma)= \sum_\alpha C_\alpha^i \psi_\alpha ({\bf R},\sigma) \, ,
\end{equation}
with $\alpha=\left\lbrace n_r, n_z,\Lambda,\Sigma\right\rbrace $.

In the present work, the mean field of the odd-$A$ nucleus is generated within the 
equal filling approximation (EFA), a prescription used in self-consistent mean-field 
calculations that preserves time-reversal invariance. In this approximation half of 
the unpaired nucleon sits in a given orbital and the other half in the time-reversed 
partner. The odd nucleon orbital, characterized by $\Omega_i=k$ and $\pi_i$, is 
chosen among those around the Fermi level, according to the experimental spin and 
parity values. The EFA should be in our view the first reasonable attempt to 
describe odd-$A$ nuclei because of its numerical advantages and of its 
reliability when comparing with the results obtained from more sophisticated approaches. 
Such a comparison has been carried out in Ref. \cite{schunck}, where the EFA was 
compared with the exact blocking procedure with time-odd mean fields fully taken into 
account. It was shown that both procedures are strictly equivalent when time-odd terms 
are neglected. Different prescriptions for the time-odd coupling constants were also 
explored in  \cite{schunck}, showing that the impact of the time-odd terms is quite 
small. The final conclusion was that the EFA is sufficiently precise for most practical 
applications. A microscopic justification of the EFA was given in Ref. \cite{robledo}, 
in terms of standard procedures of quantum statistical mechanics.

Time-odd fields vanish in the ground state of even-even nuclei, but they are different 
from zero in $I>0$ nuclei, where the time-reversal symmetry is broken. Our knowledge of 
the coupling constants of the time-odd terms is unfortunately still very limited. They 
could be constrained by experimental data on odd-$A$ nuclei, but their ability to
do so is still an open issue \cite{bender02,zalewski}.
It would be certainly interesting in the future to check the impact of going beyond 
the EFA by including explicitly odd-terms in the Skyrme interactions to be adjusted
to data on elastic magnetic electron scattering from odd-$A$ nuclei.

All the results presented in this work correspond to the Skyrme 
interaction SLy4 \cite{chabanat}, which has been thoroughly tested on many nuclear 
properties along the full nuclear chart.

The explicit expressions for all the intrinsic form factors in Eqs.(\ref{fmk}) and  
(\ref{fm2k}) in terms of these wave functions can be found in \cite{elvira1,kova}.
Expressions for the intrinsic rotational multipoles ${\cal F}^{M\lambda}_R$ can be
also found in  Ref. \cite{elvira1,elvira2} for different microscopic 
(Projected Hartree-Fock 
and cranking), as well as macroscopic (rigid rotor and irrotational flow) models.

\begin{table*}[th]
\caption{Calculated and measured charge root mean square radii $r_c$ [fm],
quadrupole deformation parameter $\beta_p$, spectroscopic nuclear electric 
quadrupole moment $Q_{\rm lab}$ [b], and magnetic moments $\mu \  [\mu_N ]$.
} 
{\begin{tabular}{cccccccccc} \hline \hline \\
Nucleus   & $I^{\pi}$ & $r_{c,{\rm th}}$  &  $r_{c,{\rm exp}}$ \cite{angeli} & $\beta_p$  
& $Q_{\rm lab, th}$ & $Q_{\rm lab, exp}$ \cite{stone} & $\mu_{\rm th, sph} $ & 
$\mu_{\rm th, def} $ & $\mu_{\rm exp} $ \cite{stone}  \\ \\
$^{17}$O   &  $5/2^+$ & 2.7429 & 2.6932(75) & -0.034    & -0.009  & -0.02578  
& -1.913  & -1.946 & -1.89379(9)   \\
$^{25}$Mg  &  $5/2^+$ & 3.0868 & 3.0284(22) &  0.338    &  0.167  & +0.199(2) 
& -1.913  & -0.864 & -0.85545(8)   \\
$^{29}$Si  &  $1/2^+$ & 3.1434 & 3.1176(52) & -0.172    &   -     &      -   
& -1.913  & -0.739 & -0.55529(3)     \\
$^{39}$K   &  $3/2^+$ & 3.4509 & 3.4349(19) &  0.026    &  0.014  & +0.0585(6) 
& +0.124  & +0.116 & +0.39157(3)  \\
$^{41}$Ca  &  $7/2^-$ & 3.4910 & 3.4780(19) & -0.020    & -0.027  & -0.090(2)/-0.066(2) 
& -1.913  & -1.936 & -1.5942(7) \\
$^{51}$V   &  $7/2^-$ & 3.6233 & 3.6002(22) & -0.027    & -0.047  & -0.043(5) 
& +5.793  & +5.817 & +5.1487057(2)   \\
$^{59}$Co  &  $7/2^-$ & 3.7891 & 3.7885(21) &  0.116    &  0.255  & +0.41(1)/+0.35(3) 
& +5.793  & +5.152 & +4.627(9) \\
$^{93}$Nb  &  $9/2^+$ & 4.3237 & 4.3240(17) & -0.042    & -0.216  & -0.37(2)/-0.32(2) 
& +6.793  & +6.691 & +6.1705(3) \\
$^{115}$In &  $9/2^+$ & 4.6175 & 4.6156(26) &  0.090    &  0.632  & +0.81(5)  
& +6.793  & +5.968 & +5.5408(2)   \\
\hline \hline
\label{table1}
\end{tabular}}
\end{table*}

\subsection{Spherical Limit}

In the spherical limit, the single-particle wave functions contain a single angular 
momentum component, so that the odd-nucleon wave function $\phi_k$ contains a single 
component $\phi_{jj}$. Thus, in this case $j=k=I_i$. 

In this limit there are no collective magnetic multipoles (${\cal F}^{M\lambda}_R=0$) 
and the intrinsic single-particle form factors are given by the following expressions,

\begin{eqnarray}
{\cal F}^{M\lambda}_{k} & = &  \langle \phi_{jj} | \hat{T} ^{M\lambda}_0 |
\phi_{jj} \rangle \nonumber \\
&& = \frac{1}{\sqrt{2j+1}} \langle jj \lambda 0 |jj \rangle 
\langle \phi_j ||  \hat{T} ^{M\lambda} ||  \phi_j \rangle   \, ,
\label{fmksph}
\end{eqnarray}
\begin{eqnarray}
{\cal F}^{M\lambda}_{2k} & = &  \langle \phi_{jj} | \hat{T} ^{M\lambda}_{2j} |
{\bar \phi}_{jj} \rangle \nonumber \\
&& = \frac{(-1)^{\lambda}}{\sqrt{2j+1}} \langle j\ -j\ \lambda \ 2j|jj \rangle 
\langle \phi_j ||  \hat{T} ^{M\lambda} ||  \phi_j \rangle   \, . 
\label{fm2ksph}
\end{eqnarray}

According to Eq. (\ref{f_rme}), for spherical nuclei one has
\begin{equation}
\left. F^{M\lambda} \right| _{\rm sph} =  \frac{1}{\sqrt{2j+1}} 
\langle \phi_j ||  \hat{T} ^{M\lambda} ||  \phi_j \rangle \, .
\end{equation}

On the other hand, using Eqs. (\ref{fmksph}) and (\ref{fm2ksph}), one can see that
$\left. F^{M\lambda} \right| _{\rm def}$   in Eq. (\ref{f_rot}) is related to 
$\left. F^{M\lambda} \right| _{\rm sph} $ by a geometrical factor $\eta _j^{\lambda}$
given by

\begin{equation}
\eta _j^{\lambda} = \langle jj\lambda 0|jj  \rangle ^2
\left[ 1+\delta_{\lambda,2j}
\frac{ \langle j\ -j\ \lambda \ 2j|jj \rangle ^2}
{ \langle  jj \lambda 0 |jj\rangle ^2} \right] \ .
\label{eta_jl}
\end{equation}
These coefficients are given explicitly by the following expressions for the two
cases $\lambda < 2j$ and $\lambda = 2j$:

\begin{equation}
\eta _j^{\lambda < 2j} = \frac{(2j+1)! (2j)!}{(2j+\lambda +1)! (2j-\lambda)!} \, ,
\end{equation}

\begin{equation}
\eta _j^{\lambda = 2j} = \frac{2j+1}{4j+1}\left[ 1+\frac{[(2j)!]^2}{(4j)!} \right] \, .
\end{equation}

In the next section, we denote by  $\left. F^{M\lambda} \right| _{\rm sph\ limit}$ the form 
factor obtained with the deformed codes for the case of odd-$A$ spherical nuclei using 
the above equations,

\begin{equation}
\left. F^{M\lambda} \right| _{\rm sph\ limit} =
[\eta _j^{\lambda}]^{-1} \left. F^{M\lambda} \right| _{\rm def} \ ,
\label{sphlimit}
\end{equation}
with  $\left. F^{M\lambda} \right| _{\rm def}$  as in Eq. (\ref{f_rot}) and $\eta _j^{\lambda}$ 
as in Eq. (\ref{eta_jl}). These calculations are applied not only to the spherical 
nuclei ($^{17}$O, $^{39}$K, $^{41}$Ca  $^{51}$V, and $^{93}$Nb), but also to the deformed 
nuclei ($^{25}$Mg, $^{29}$Si, $^{59}$Co, and $^{115}$In) in the limit of zero deformation, 
using constrained HF+BCS solutions for $\beta=0$. 

The geometrical factors  $\eta _j^{\lambda}$ in Eq. (\ref{sphlimit}) are due to the loss 
of a favored intrinsic direction that takes place when one goes from the deformed to 
the spherical limit.

\subsection{Natural Orbitals and Nucleon Correlation Effects on Electron Elastic 
Magnetic Scattering from Nuclei}
\label{subsec:no}

It is known (see, e.g., \cite{AHP,Kadrev96}) that many experimental data of scattering 
and reactions on nuclei show the existence of sizable nucleon-nucleon (NN) correlation 
effects on nuclear properties that cannot be described correctly within the mean-field 
approximation in the nuclear theory. This concerns mainly the form of the single-particle
wave functions, the occupation probabilities, the spectral functions, the momentum 
distributions of nucleons and clusters, and others. It became possible to restore the 
single-particle picture in the methods in which the NN correlations are accounted for 
by using the so-called natural orbitals (NOs) and natural occupation probabilities 
within the NO representation \cite{Lowdin55} of the one-body density matrix (OBDM) that 
corresponds to the correlated ground state of the system. The NOs and the natural 
occupation numbers for the nucleus with $A$ particles can be obtained by the 
diagonalization of the OBDM in a given model solving the equation:

\begin{equation}
\int d{\bf r^{\prime}}\rho({\bf r},{\bf r^{\prime}})\psi_{\alpha}({\bf r^{\prime}})=
n_{\alpha}\psi_{\alpha}({\bf r}),
\label{eq:diagonal}
\end{equation}
where $\psi_{\alpha}({\bf r})$ are the NOs and $n_{\alpha}$ are the natural occupation 
numbers that fulfill the conditions:
\begin{equation}
0 \leq n_{\alpha} \leq 1 ; \;\;\;\;\; \sum_{\alpha}n_{\alpha}=A \;.
\label{eq:conditions}
\end{equation}
Thus, in the NO representation \cite{Lowdin55} the OBDM has the form:
\begin{equation}
\rho({\bf r},{\bf r^{\prime}})=\sum_{\alpha}n_{\alpha}
\psi_{\alpha}^{*}({\bf r^{\prime}})\psi_{\alpha}({\bf r}).
\label{eq:obdm}
\end{equation}
In the present work, the NOs related to the OBDM in the Coherent Density Fluctuation 
Model (CDFM) \cite{Antonov79,AHP,Antonov89,Antonov94,Kadrev96} are obtained for 
different nuclei and used in the calculations of the magnetic form factors.

The CDFM is based \cite{AHP,Antonov79} on the $\delta$-function limit of the 
generator-coordinate method \cite{Griffin57}. The OBDM in the model has the form:
\begin{equation}
\rho({\bf r},{\bf r^{\prime}})=\int |{\cal F}(x)|^{2}
\rho_{x}({\bf r},{\bf r^{\prime}})dx,
\label{eq:cdfm}
\end{equation}
being an infinite superposition of the OBDMs $\rho_{x}({\bf r},{\bf r^{\prime}})$ related 
to the one in the plane-wave case:
\begin{eqnarray}
\rho_{x}({\bf r},{\bf r^{\prime}})&=&3\rho_{0}(x) \frac{j_{1}(k_{F}(x)|{\bf r}-
{\bf r^{\prime}}|)}{k_{F}(x)|{\bf r}-{\bf r^{\prime}}|}\nonumber \\  
& \times & \Theta \left (x-\frac{|{\bf r}+{\bf r^{\prime}}|}{2}\right ) \, ,
\label{eq:flucton}
\end{eqnarray}
with

\begin{equation}
\rho_{0}(x)=\frac{3A}{4\pi x^{3}}  \, ,
\label{eq:rho0}
\end{equation}

\begin{equation}
k_{F}(x)=\left(\frac{3\pi^{2}}{2}\rho_{0}(x)\right )^{1/3}\equiv \frac{\alpha}{x}, \;\; 
\alpha=\left(\frac{9\pi A}{8}\right )^{1/3}. 
\label{eq:fermi}
\end{equation}
The weight function $|{\cal F}(x)|^{2}$ in (\ref{eq:cdfm}) in the case of 
monotonically-decreasing density distributions ($d\rho(r)/dr\leq 0$) can be obtained 
from the density:
\begin{equation}
|{\cal F}(x)|^{2}=-\frac{1}{\rho_{0}(x)} \left. \frac{d\rho(r)}{dr}\right |_{r=x} .
\label{eq:weight}
\end{equation}
The applications of the NOs to studies of the magnetic form factors are a part of 
general studies of the correlated OBDMs and their usage to analyses of nuclear 
states and processes.

%%%%%%%%%%%%%%%%%%%%%%%%%%%%Fig1%%%%%%%%%%%%%%%%%%%%%%%%%%%%%%%%%%%%%%%%%%%%%%%%%%%%%%%%
\begin{figure}[htb]
\centering
\includegraphics[width=70mm]{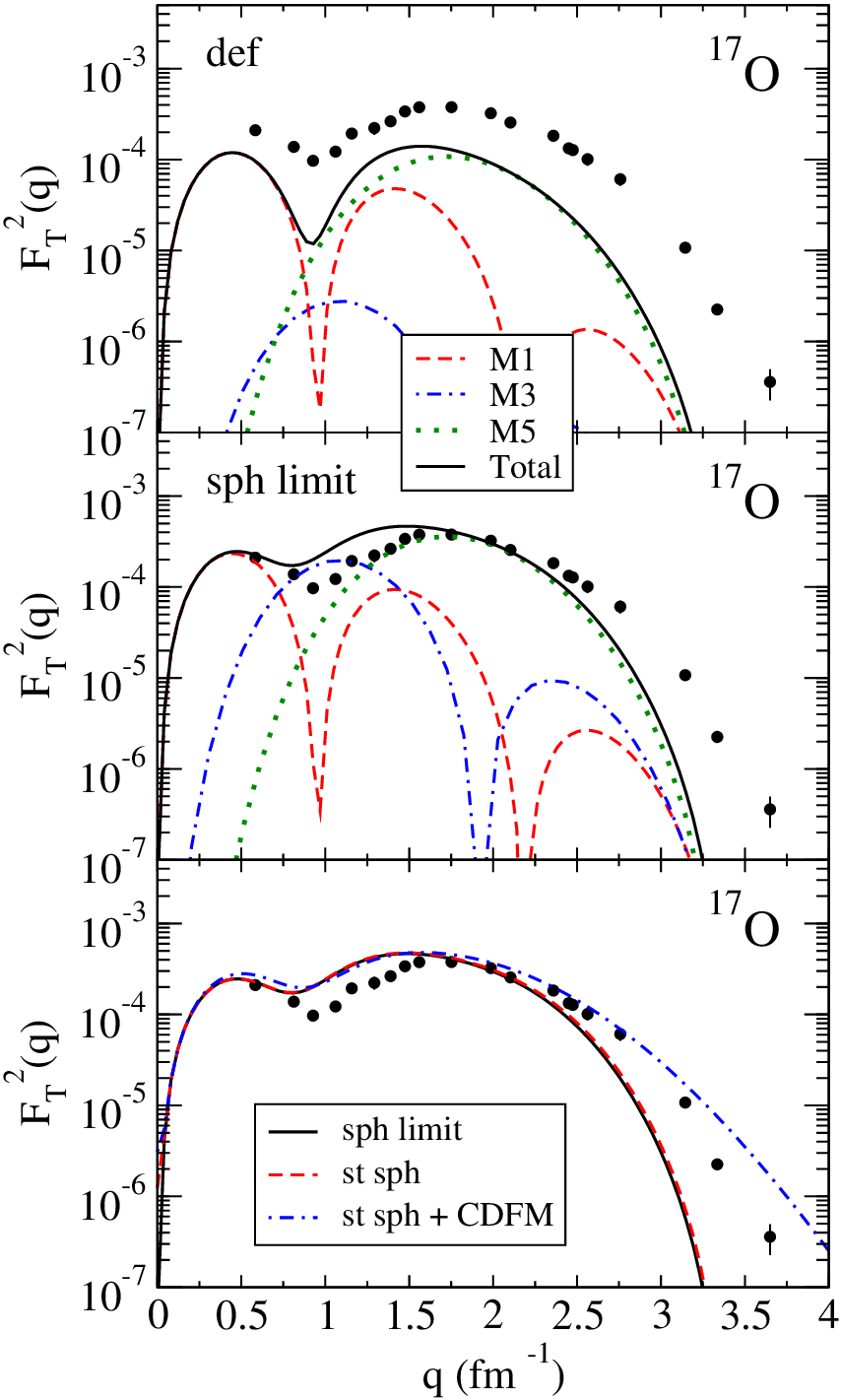}
\caption{Magnetic form factors of $^{17}$O ($I^{\pi}=5/2^+$) decomposed into M1, M3, 
and M5 multipole components in the deformed model (top) and in its spherical limit 
(middle). We also show (bottom) the results from standard spherical calculations 
alone and together with CDFM (see text). Data are taken from \cite{donnelly,hynes}.
}
\label{fig_17o_ff}
\end{figure}
%%%%%%%%%%%%%%%%%%%%%%%%%%%%%%%%%%%%%%%%%%%%%%%%%%%%%%%%%%%%%%%%%%%%%%%%%%%%%%%%%%%%%%%%%

%%%%%%%%%%%%%%%%%%%%%%%%%%%%Fig2%%%%%%%%%%%%%%%%%%%%%%%%%%%%%%%%%%%%%%%%%%%%%%%%%%%%%%%%
\begin{figure}[htb]
\centering
\includegraphics[width=70mm]{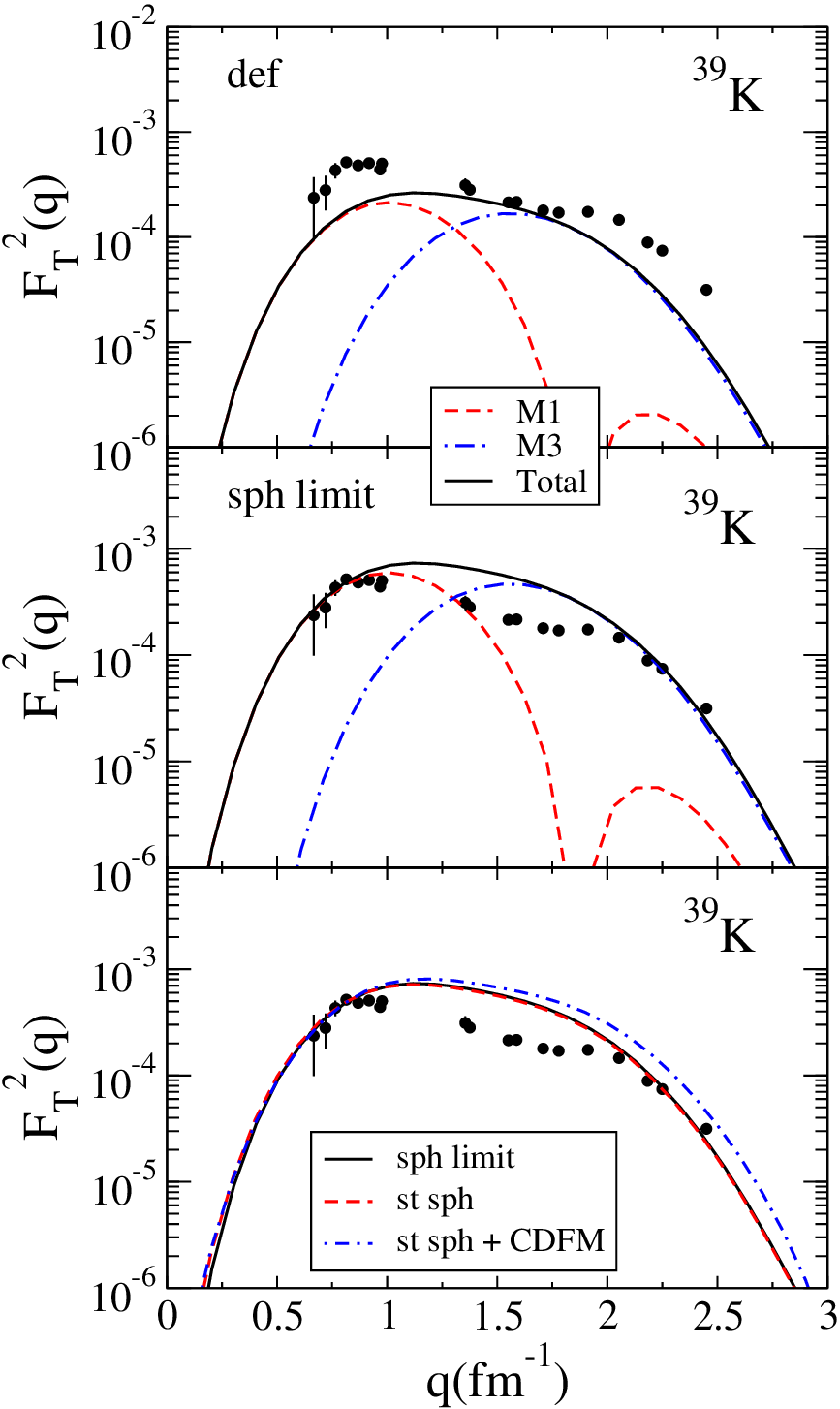}
\caption{Same as in Fig. \ref{fig_17o_ff}, but for $^{39}$K ($I^{\pi}=3/2^+$) decomposed 
into M1 and M3 multipole components. Data are taken from \cite{donnelly} and references 
therein.
}
\label{fig_39k_ff}
\end{figure}
%%%%%%%%%%%%%%%%%%%%%%%%%%%%%%%%%%%%%%%%%%%%%%%%%%%%%%%%%%%%%%%%%%%%%%%%%%%%%%%%%%%%%%%%%

%%%%%%%%%%%%%%%%%%%%%%%%%%%%Fig3%%%%%%%%%%%%%%%%%%%%%%%%%%%%%%%%%%%%%%%%%%%%%%%%%%%%%%%%
\begin{figure}[htb]
\centering
\includegraphics[width=70mm]{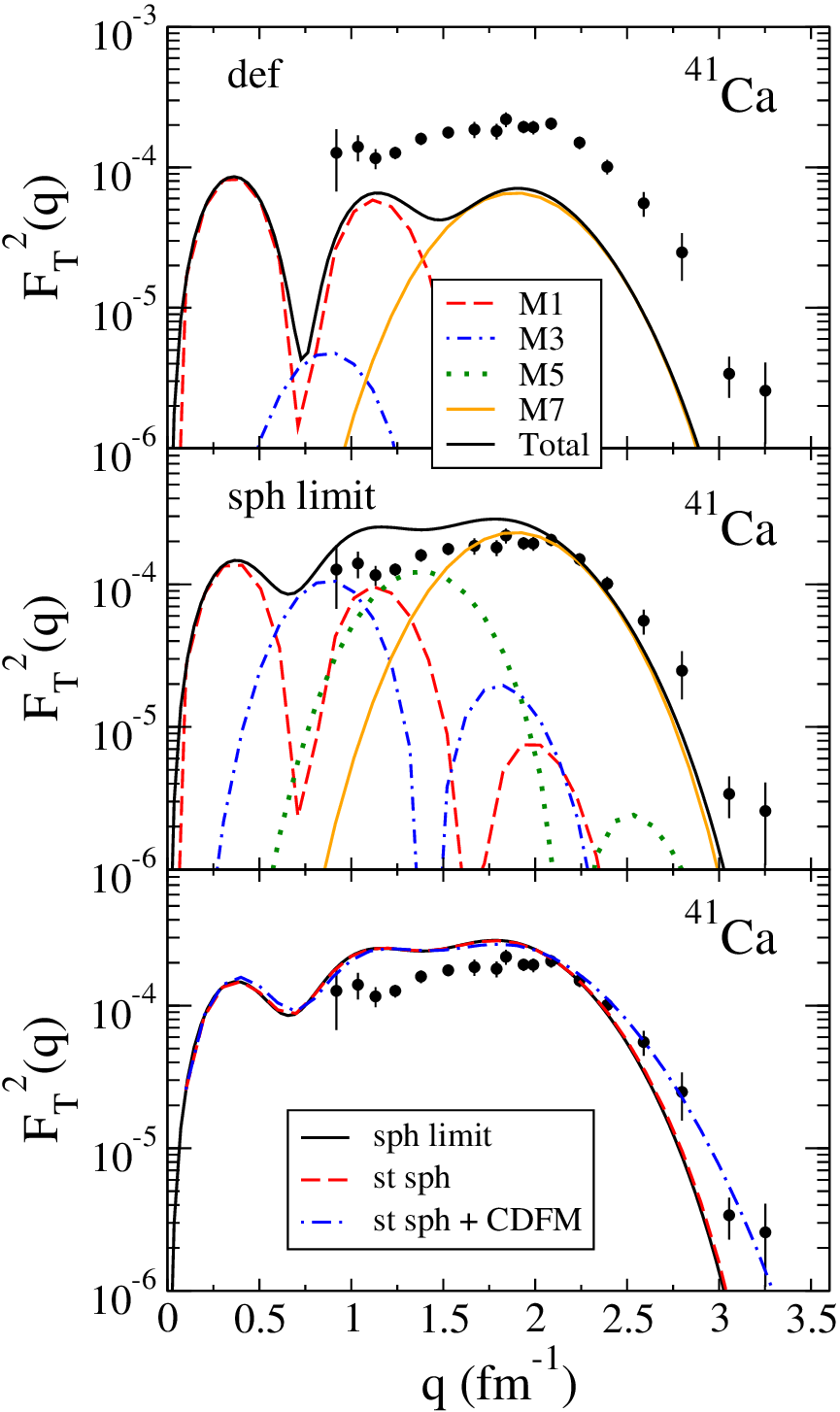}
\caption{Same as in Fig. \ref{fig_17o_ff}, but for $^{41}$Ca ($I^{\pi}=7/2^-$) decomposed 
into M1, M3, M5, and M7 multipole components. Data are taken from \cite{baghaei}.
}
\label{fig_41ca_ff}
\end{figure}
%%%%%%%%%%%%%%%%%%%%%%%%%%%%%%%%%%%%%%%%%%%%%%%%%%%%%%%%%%%%%%%%%%%%%%%%%%%%%%%%%%%%%%%%%

%%%%%%%%%%%%%%%%%%%%%%%%%%%%Fig4%%%%%%%%%%%%%%%%%%%%%%%%%%%%%%%%%%%%%%%%%%%%%%%%%%%%%%%%
\begin{figure}[htb]
\centering
\includegraphics[width=70mm]{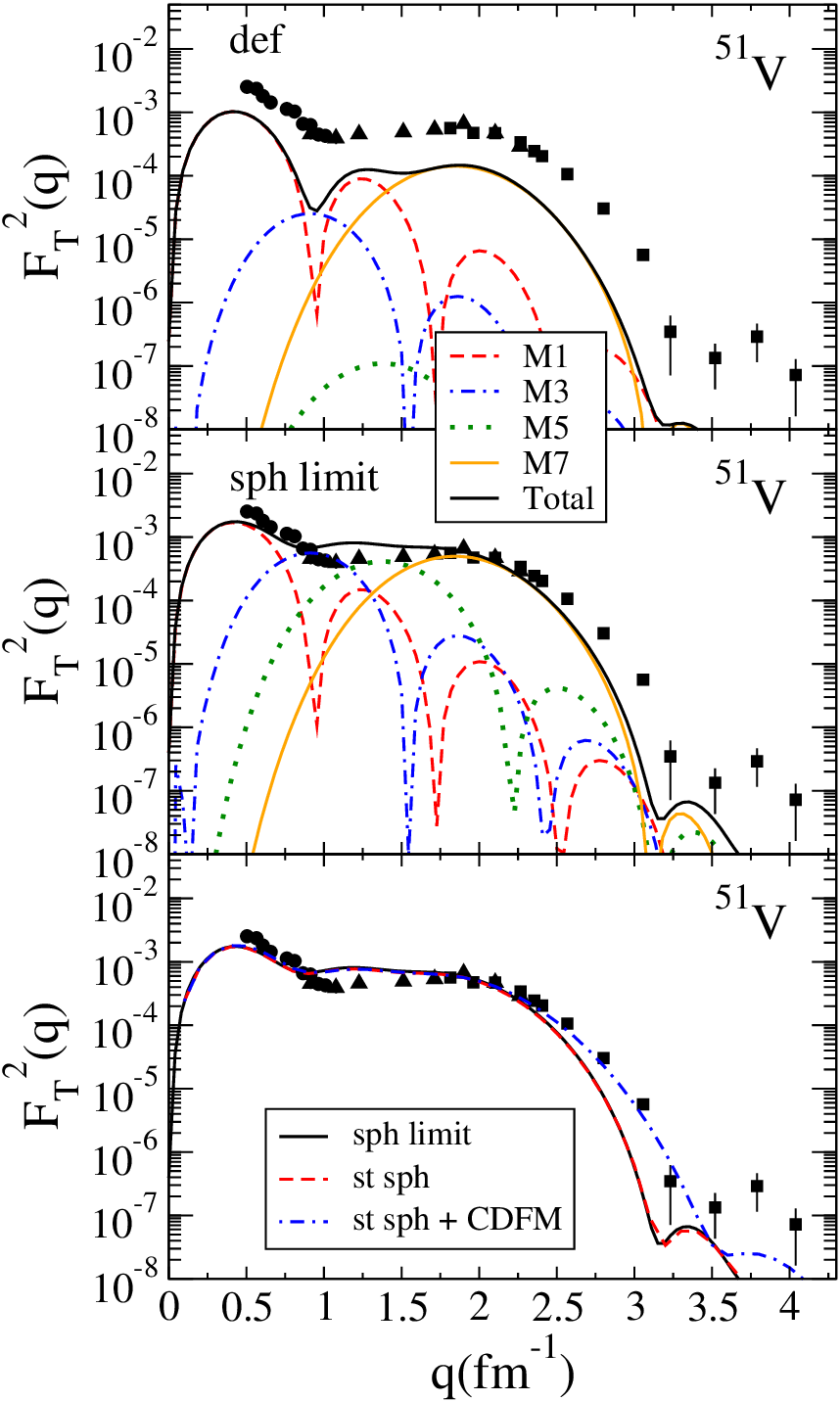}
\caption{Same as in Fig. \ref{fig_17o_ff}, but for $^{51}$V ($I^{\pi}=7/2^-$) 
decomposed into M1, M3, M5, and M7 multipole components. Data are taken 
from Refs. \cite{donnelly} (circles), \cite{arita81} (triangles) and 
\cite{platchkov83} (squares).
}
\label{fig_51v_ff}
\end{figure}
%%%%%%%%%%%%%%%%%%%%%%%%%%%%%%%%%%%%%%%%%%%%%%%%%%%%%%%%%%%%%%%%%%%%%%%%%%%%%%%%%%%%%%%%%

%%%%%%%%%%%%%%%%%%%%%%%%%%%%Fig5%%%%%%%%%%%%%%%%%%%%%%%%%%%%%%%%%%%%%%%%%%%%%%%%%%%%%%%%
\begin{figure}[htb]
\centering
\includegraphics[width=70mm]{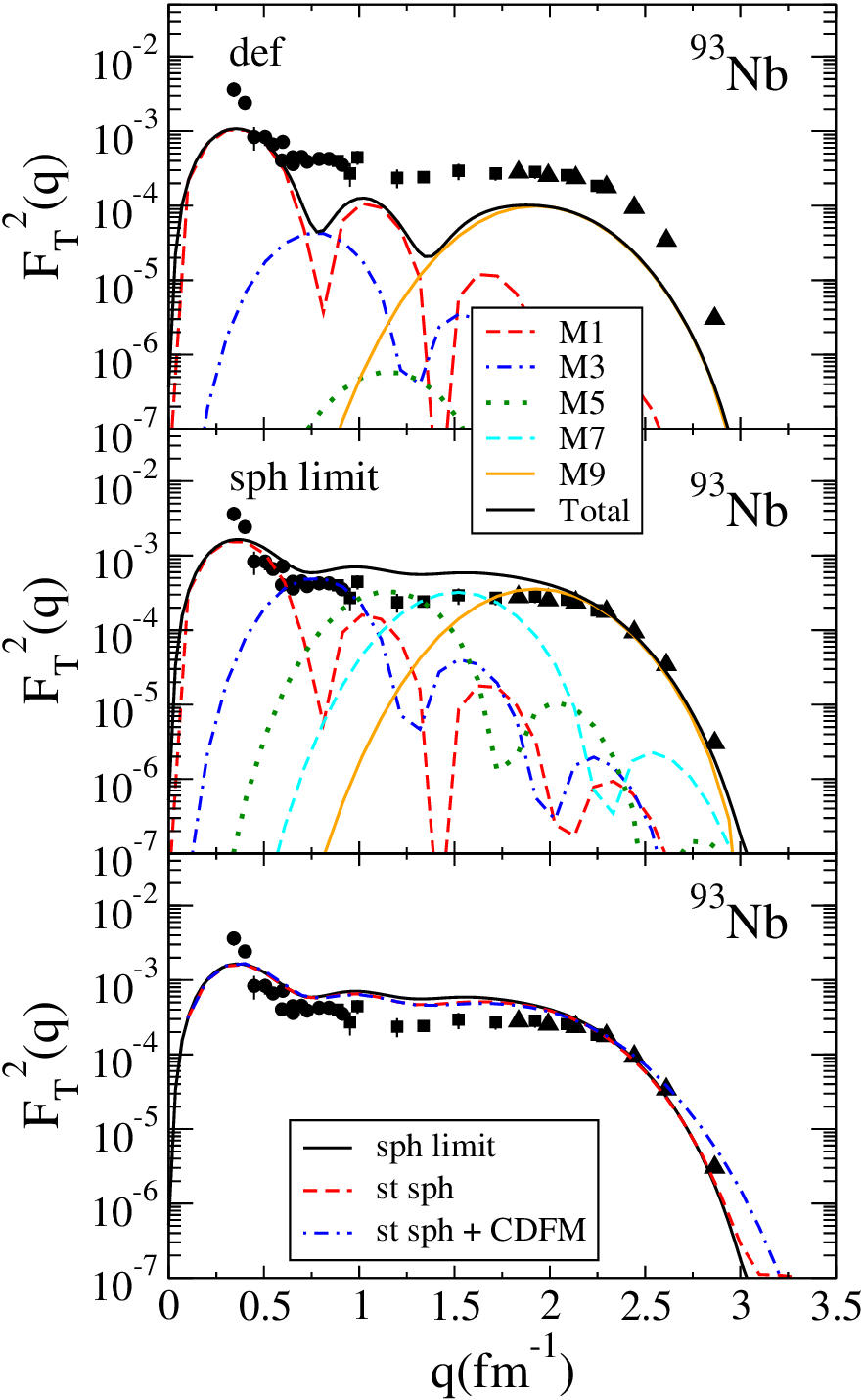}
\caption{Same as in Fig. \ref{fig_17o_ff}, but for $^{93}$Nb ($I^{\pi}=9/2^+$) 
decomposed into M1, M3, M5, M7, and M9 multipole components. Data are taken 
from \cite{donnelly} (circles), \cite{york} (squares), and \cite{platchkov82}
(triangles).
}
\label{fig_93nb_ff}
\end{figure}
%%%%%%%%%%%%%%%%%%%%%%%%%%%%%%%%%%%%%%%%%%%%%%%%%%%%%%%%%%%%%%%%%%%%%%%%%%%%%%%%%%%%%%%%%

%%%%%%%%%%%%%%%%%%%%%%%%%%%%Fig6%%%%%%%%%%%%%%%%%%%%%%%%%%%%%%%%%%%%%%%%%%%%%%%%%%%%%%%%
\begin{figure}[htb]
\centering
\includegraphics[width=70mm]{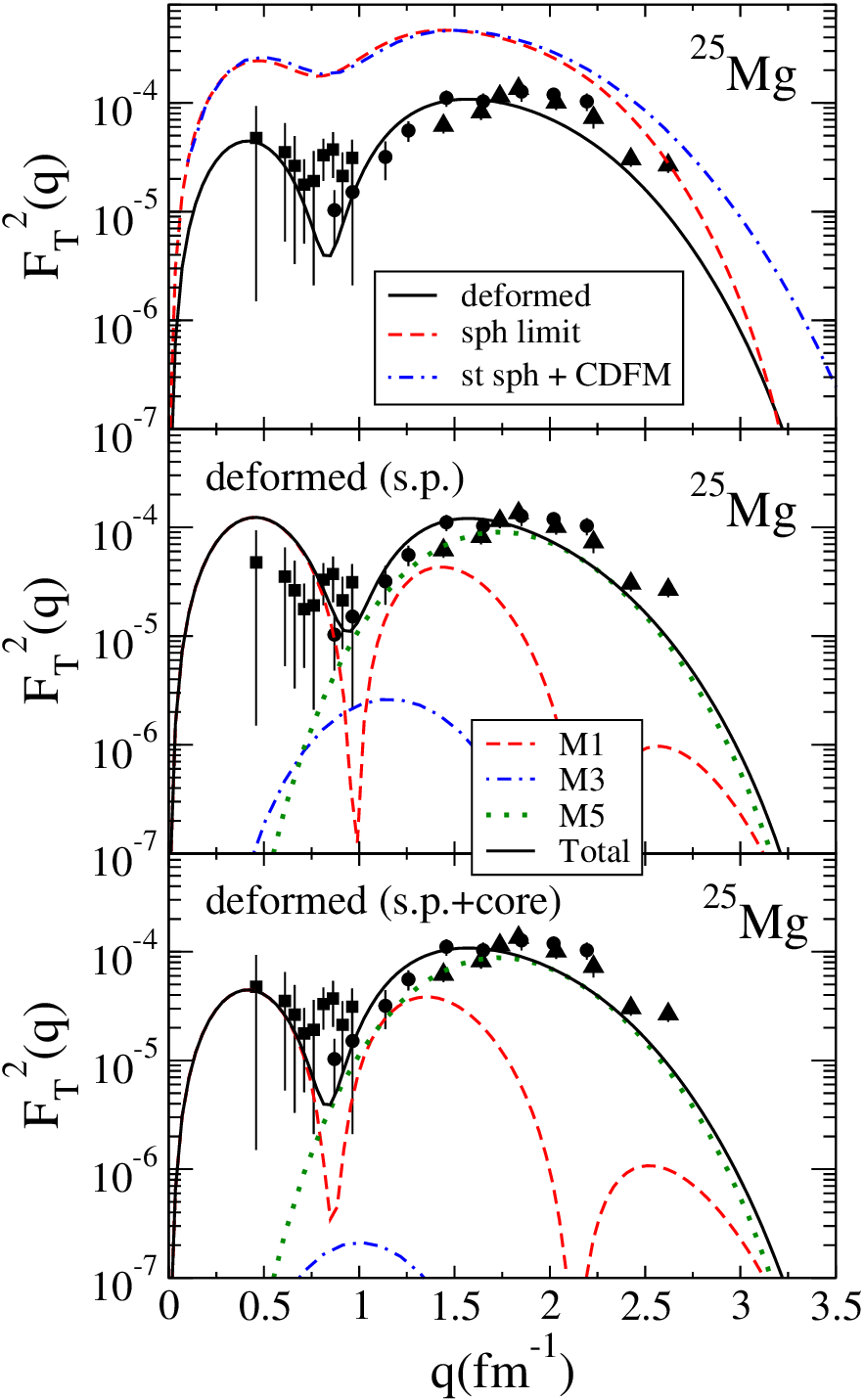}
\caption{Magnetic form factors of $^{25}$Mg ($I^{\pi}=5/2^+$) in the deformed formalism
and in its spherical limit, as well as in the spherical CDFM (top). The multipole 
decomposition is shown in the deformed model for single-particle contributions (middle) 
and including collective contributions from the rotating core in the cranking 
approximation (bottom). Data used in \cite{donnelly} are taken from \cite{york} (circles)
and \cite{euteneuer} at different kinematical conditions (squares and triangles).
}
\label{fig_25mg_ff_sph_def}
\end{figure}
%%%%%%%%%%%%%%%%%%%%%%%%%%%%%%%%%%%%%%%%%%%%%%%%%%%%%%%%%%%%%%%%%%%%%%%%%%%%%%%%%%%%%%%%%

%%%%%%%%%%%%%%%%%%%%%%%%%%%%Fig7%%%%%%%%%%%%%%%%%%%%%%%%%%%%%%%%%%%%%%%%%%%%%%%%%%%%%%%%
\begin{figure}[htb]
\centering
\includegraphics[width=70mm]{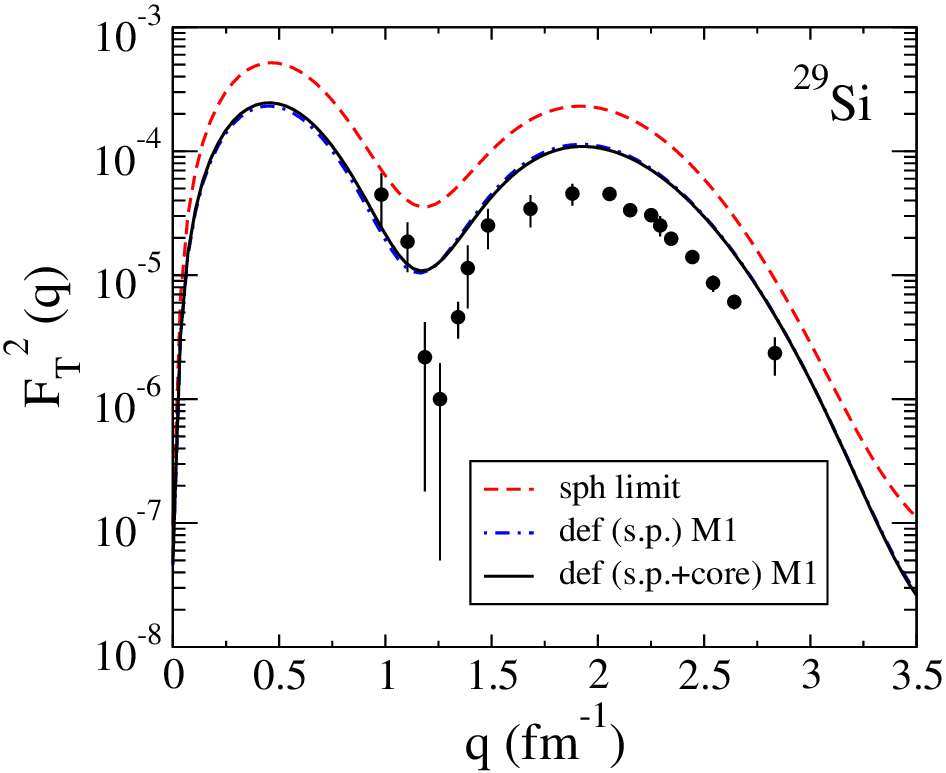}
\caption{Magnetic form factors of $^{29}$Si ($I^{\pi}=1/2^+$) in the deformed and 
spherical models. Only the M1 multipole contributes. Data are taken from \cite{donnelly}
and references therein.
}
\label{fig_29si_ff_sph_def}
\end{figure}
%%%%%%%%%%%%%%%%%%%%%%%%%%%%%%%%%%%%%%%%%%%%%%%%%%%%%%%%%%%%%%%%%%%%%%%%%%%%%%%%%%%%%%%%%

%%%%%%%%%%%%%%%%%%%%%%%%%%%%Fig8%%%%%%%%%%%%%%%%%%%%%%%%%%%%%%%%%%%%%%%%%%%%%%%%%%%%%%%%
\begin{figure}[htb]
\centering
\includegraphics[width=70mm]{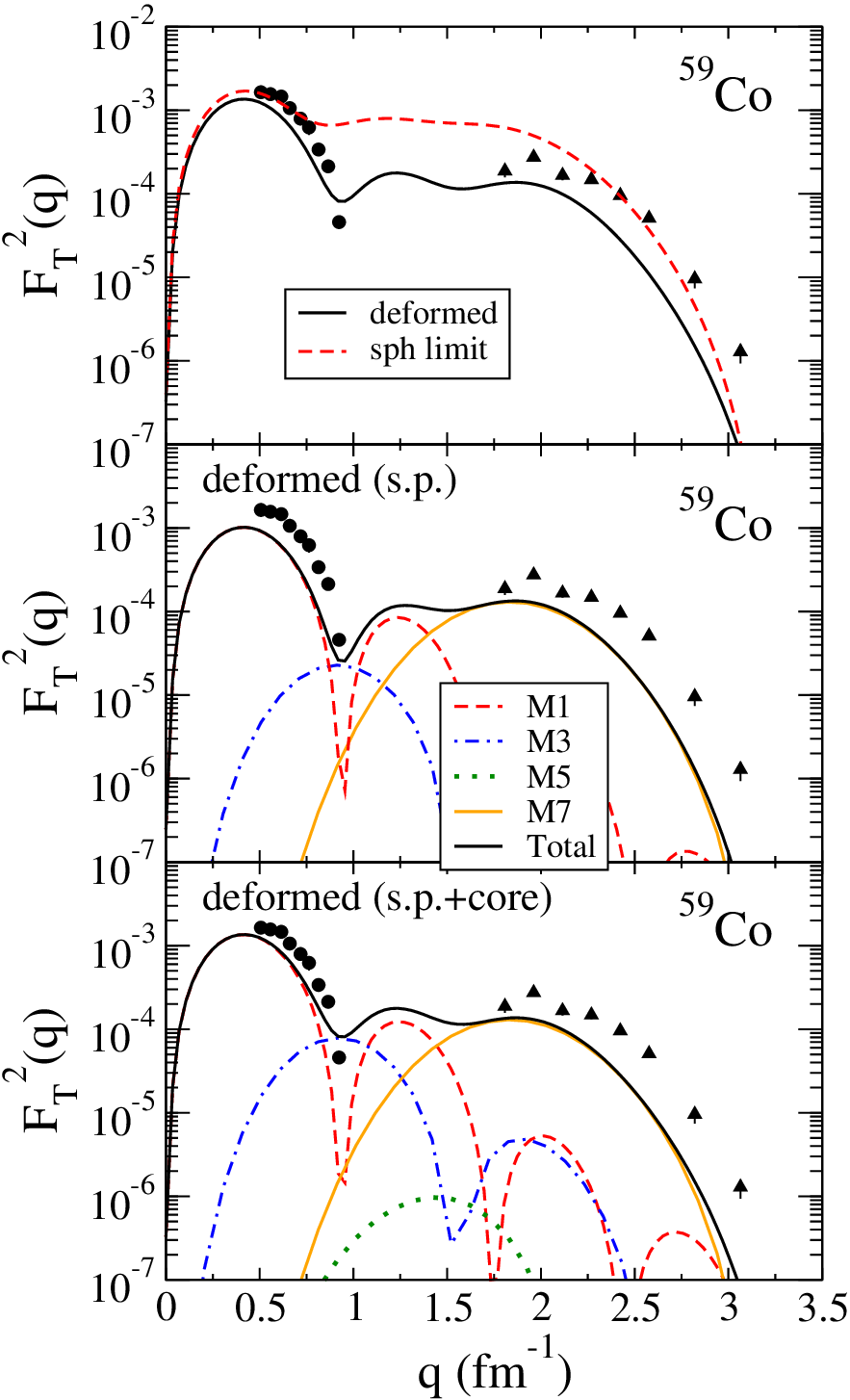}
\caption{Magnetic form factors of $^{59}$Co ($I^{\pi}=7/2^-$) in the deformed and 
spherical models (top). The multipole decomposition is shown in the deformed model
for single-particle contributions (middle) and including collective contributions 
from the rotating core in the cranking approximation (bottom). 
Data used in \cite{donnelly} are taken from \cite{devries70} (circles) and 
\cite{platchkov82} (triangles).
}
\label{fig_59co_ff_sph_def}
\end{figure}
%%%%%%%%%%%%%%%%%%%%%%%%%%%%%%%%%%%%%%%%%%%%%%%%%%%%%%%%%%%%%%%%%%%%%%%%%%%%%%%%%%%%%%%%%

%\clearpage

%%%%%%%%%%%%%%%%%%%%%%%%%%%%Fig9%%%%%%%%%%%%%%%%%%%%%%%%%%%%%%%%%%%%%%%%%%%%%%%%%%%%%%%%
\begin{figure}[htb]
\centering
\includegraphics[width=70mm]{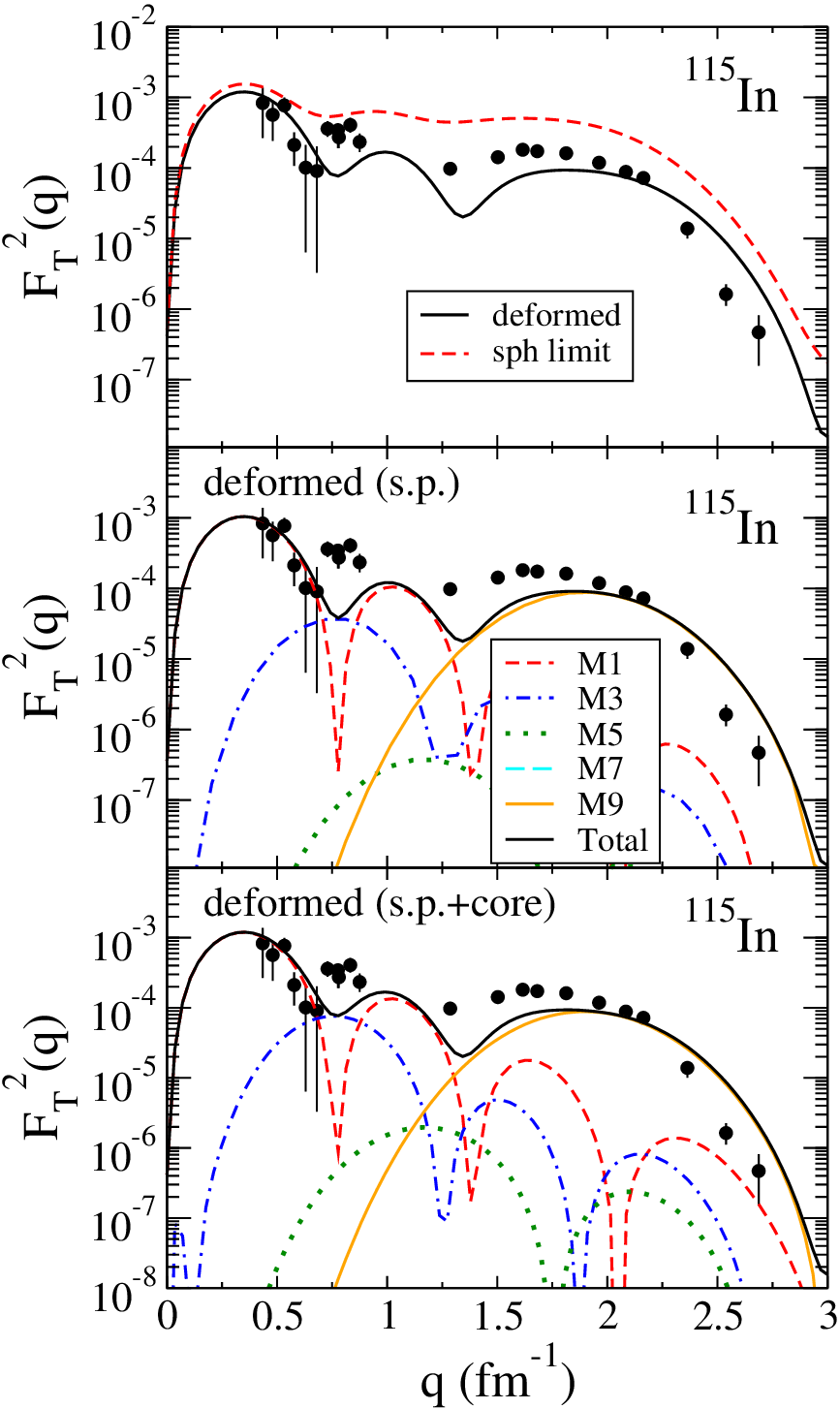}
\caption{Same as in Fig. \ref{fig_59co_ff_sph_def}, but for $^{115}$In ($I^{\pi}=9/2^+$) 
Data are taken from \cite{donnelly} and references therein.
}
\label{fig_115in_ff_sph_def}
\end{figure}
%%%%%%%%%%%%%%%%%%%%%%%%%%%%%%%%%%%%%%%%%%%%%%%%%%%%%%%%%%%%%%%%%%%%%%%%%%%%%%%%%%%%%%%%%

\section{Results and Discussion }

The nuclei studied here have been chosen because experimental information on elastic
magnetic electron scattering form factors is available and a comparison with the 
calculated ones can be performed to test the reliability of the theoretical models.

Table \ref{table1} shows ground state properties of the nuclei under study, namely,
charge root-mean-square radii $r_c$, spectroscopic  electric quadrupole moments
$Q_{\rm lab}$, and magnetic moments $\mu$. The results are compared with experimental 
data from \cite{angeli} for radii and from \cite{stone} for quadrupole and magnetic
moments.

The relationship between the intrinsic quadrupole moment $Q_0$ and the quadrupole
deformation parameter $\beta_p$ is given by
\begin{equation}
Q_0 = \sqrt{\frac{5}{\pi}} Ze \langle r^2 \rangle \beta_p \, ,
\end{equation}
where $<r^2>$ is the nuclear mean-square radius. The measured quadrupole moment 
$Q_{\rm lab}$ is related to the intrinsic quadrupole moment $Q_0$ by
\begin{equation}
Q_{\rm lab} = \frac{3k^2 - I(I+1)}{(I+1)(2I+3)} e Q_0 \, .
\end{equation}
The measured electric quadrupole moments in Table \ref{table1} correspond to 
ground states $I=k$. Of course, $Q_{\rm lab} = 0$ for $I=k=1/2$.

The magnetic moments in the deformed case $\mu_{\rm th, def}$ are obtained from the 
expression,

\begin{eqnarray}
\mu_I &=& g_R I + \frac{k^2}{I+1} \big[ g_k - g_R +  \nonumber \\
&& \delta_{k,1/2} (2I+1)(-1)^{I+1/2} \sqrt{2} g_{2k} \big] \, ,
\end{eqnarray}
where $g_R$, $g_k$, and $g_{2k}$ are defined in Ref. \cite{elvira1} and calculated in the 
cranking approximation. We also show for comparison the Schmidt values $\mu_{\rm th, sph}$ 
obtained in the spherical limit.

The agreement between the calculated and the measured $r_c$ is quite good, especially
in the medium-mass nuclei from  $^{29}$Si to  $^{115}$In, where the error is less than 
1\%. This error is slightly larger in the lighter nuclei. The quadrupole deformation 
parameters $\beta_p$ in Table \ref{table1} show that the isotopes $^{17}$O, $^{39}$K, 
$^{41}$Ca, $^{51}$V, and $^{93}$Nb could be treated as spherical, while $^{25}$Mg, 
$^{29}$Si, $^{59}$Co, and $^{115}$In, should be treated as deformed. The calculated and 
measured $Q_{\rm lab}$ agree in sign, as well as in magnitude in most cases.

Concerning magnetic moments, for the spherical nuclei ($^{17}$O, $^{39}$K, $^{41}$Ca, 
$^{51}$V, and $^{93}$Nb), both the Schmidt and the deformed values show a reasonable 
agreement with the experimental data, except in the case of $^{39}$K, where the 
calculations overestimate the data in both cases. On the other hand, for the deformed 
nuclei ($^{25}$Mg, $^{29}$Si, $^{59}$Co, and $^{115}$In), there is a clear improvement
of the agreement with experiment in the deformed formalism.

The reason why deformation does not improve the results on spherical nuclei is simply 
that this is not a proper model to describe them. For odd-$A$ nuclei, the deformed model 
we use is valid for well deformed nuclei where the coupling between the unpaired 
particle and the rotation of the even-even core can be treated in the rotor plus 
quasiparticle approximation (using Skyrme HF+BCS single-particle states). The spherical 
model for odd-$A$ nuclei assumes that the ground state spin $I$ and parity $\pi$ are 
those of the single nucleon determined from the quantum numbers of the last filled 
spherical orbital, $I=j$, $\pi=(-1)^\ell$. Typically, the deformed model can be applied 
when the quadrupole deformation parameter $\beta$ is larger than 0.1, whereas the 
spherical model is valid for zero or very small deformations. Nuclei with small 
intrinsic deformations, such as those in Table I ($^{17}$O, $\beta=-0.034$;  $^{39}$K, 
$\beta=0.026$; $^{41}$Ca, $\beta=-0.020$; $^{51}$V, $\beta=-0.027$; $^{93}$Nb, 
$\beta=-0.042$) can be considered to be spherical or rather soft nuclei, 
where the deformed formalism does not imply necessarily an improvement over the spherical 
formalism. It is worth mentioning that in the case of $^{93}$Nb, where the deformed 
formalism improves somewhat the magnetic moment, the deformation is larger than in the 
other nuclei considered as spherical. The description of these soft nuclei could be 
refined with a more sophisticated treatment of the coupling of single particle and 
collective aspects, but this is beyond the purpose of the present study. In any case, 
in the soft nuclei, the corrections introduced by 
deformation are rather small and do not change significantly the spherical results.

\subsection{Spherical nuclei}

In this subsection we study the magnetic form factors in several nuclei that according
to Table \ref{table1} can be taken as spherical, namely,  $^{17}$O, $^{39}$K, $^{41}$Ca,  
$^{51}$V, and $^{93}$Nb.

In Figs. \ref{fig_17o_ff}-\ref{fig_93nb_ff} we show the calculations from the deformed
model in the top panels. The middle panels contain the results in the spherical limit
defined by Eq. (\ref{sphlimit}). The geometrical factors $\eta_j^{\lambda}$ relating
the spherical and the deformed form factors in Eq. (\ref{eta_jl}) are given explicitly 
in Table \ref{table_eta} for each $j$ state and $\lambda$ multipole. It is worth noting 
that, in the spherical limit, there are no collective contributions ${\cal F}^{M\lambda}_R$ 
from rotations of the core. The bottom panels show a comparison between the results in 
the spherical limit (sph limit) with the results obtained from a standard spherical 
calculation (st sph) based on the code by H. P. Blok and J. H. Heisenberg \cite{blok}. 
This comparison helps us to check the reliability of the spherical limit. 
We also test the 
role of NN correlations (st sph + CDFM) by incorporating these effects and comparing 
the results. Correlation methods beyond the 
mean-field picture are needed, especially to account for the behavior at high momentum 
transfer. The NO representation is a convenient way to deal with these correlations.
In Ref. \cite{Kadrev96} NOs, obtained within the CDFM, were used to calculate 
electron magnetic scattering form factors in $^{17}$O and $^{41}$Ca, which are
examples of nuclei with a single nucleon outside a doubly-closed core. In this work
we extend those calculations to other spherical nuclei.

\begin{table*}[th]
\caption{Geometrical factors ($\eta_j^{\lambda}$ in Eq. (\ref{eta_jl})) for $j$ states and
$\lambda$ multipoles.} 
{\begin{tabular}{ccccccccccccccccccccc} \hline \hline \\
$j$ && 1/2 && \multicolumn{2}{c}{3/2} && \multicolumn{3}{c}{5/2} && 
\multicolumn{4}{c}{7/2} && \multicolumn{5}{c}{9/2} \\
\cline{5-6} \cline{8-10} \cline {12-15} \cline{17-21} 
$\lambda$ && 1 && 1 & 3 && 1 & 3 & 5 && 1 & 3 & 5 & 7 && 1 & 3 & 5 & 7 & 9 \\ \\
$\eta_j^{\lambda}$  && 1 && 0.60 & 0.60 && 0.7143 & 0.1190 & 0.5476 && 0.7778 & 0.2121 & 
0.0163 & 0.5335 &&  0.8182 & 0.2937 & 0.0419 & 0.0019 & 0.5263 \\
\hline \hline
\label{table_eta}
\end{tabular}}
\end{table*}

Figure \ref{fig_17o_ff} for $^{17}$O ($I^{\pi}=5/2^+$) shows the first example of this 
comparison. Data are from \cite{donnelly,hynes}. In the deformed case, contributions 
from the M3 multipole are negligible due to the reduction factor $\eta_j^{\lambda} = 0.119$,
but in the spherical case it makes a difference by filling the region between the two 
peaks. The first peak is totally due to the M1 multipole, while the second peak and the 
tail at large $q$ is determined by the M5 multipole. The spherical description clearly 
improves the agreement with experiment due to the geometrical factors that scale the 
various multipoles. The agreement between the results from the 
spherical limit and the results from \cite{blok} is remarkable, which is the general 
trend observed in all cases studied in this subsection. The main effect of the NN 
correlations calculated with natural orbits within the CDFM \cite{Kadrev96} is to modify 
the tails by shifting the form factors to higher $q$-values, thus improving the agreement 
with experiment. This feature is also observed in all cases studied. 

Figure \ref{fig_39k_ff} shows the same results, but for $^{39}$K ($I^{\pi}=3/2^+$). Now, 
M1 (M3) determines the first (second) peak. The enhancement of the multipoles in the 
spherical limit improves somewhat the agreement with experimental data. The effect of 
the NN correlations is similar to the previous case. There is a difficulty in the
description of the region $1<q<2$ fm$^{-1}$, where the data are overestimated by the
calculations. This is the region where the M3 multipole is dominant.

In Fig.  \ref{fig_41ca_ff} we show the results for $^{41}$Ca ($I^{\pi}=7/2^-$). In the 
deformed case we get a three peaked profile. The two first peaks are due to the M1 
multipole, while the third peak is determined by M7. M3 and M5 represent a negligible 
contribution, partly due to their geometrical factors (see Table \ref{table_eta}).
In the spherical limit all the multipoles, but especially M3 and M5, are enhanced
with respect to the deformed case, producing a much better agreement with experiment. 
In particular, M3 and M5 fill the total form factor in the range between $0.5<q<2$ 
fm$^{-1}$, getting the results closer to the data from Ref. \cite{baghaei}, although 
slightly overestimating them in the region $1<q<2$ fm$^{-1}$. The tails are again 
improved when NN correlations are included.

Figure \ref{fig_51v_ff} shows the results for $^{51}$V ($I^{\pi}=7/2^-$) with data
from \cite{donnelly,arita81,platchkov83}. Again, M1 determines the first peak and 
M7 the broad second peak and the tail. The spherical limit with the multipoles
enhanced reproduces better the data. The NN correlations improve the decay of the tail 
at large momentum transfer.

Finally,  Fig.  \ref{fig_93nb_ff} for $^{93}$Nb ($I^{\pi}=9/2^+$) shows that, in the 
deformed case, M1 is responsible for the two first peaks in the form factor and M9 
determines the third peak and the structure of the tail at high momentum transfer. 
The role of M3, M5, and M7 is irrelevant due to the reduction factors. In the spherical 
limit, all the multipoles play a role. M1 determines the region below $q=0.5$ fm$^{-1}$, 
M3 between $q=0.5$ and $q=1$ fm$^{-1}$, M5 between $q=1.0$ and $q=1.5$ fm$^{-1}$, M7 
between $q=1.5$ and $q=2.0$ fm$^{-1}$, and finally M9 the behavior beyond $q=2.0$ 
fm$^{-1}$. The final result is a smeared profile in much better agreement with the 
experiment \cite{donnelly,york,platchkov82}. The NN correlations improve slightly 
the agreement with experiment.

In summary, we have found that in the case of the odd-$A$ spherical nuclei studied, 
the spherical formalism reproduces quite reasonably the main features of the elastic
magnetic form factors measured. The results from standard spherical formalism perfectly 
agree with the results obtained in the spherical limit of the deformed formalism.
The latter is shown to be inadequate to describe spherical or soft
nuclei with very small deformations, unless we previously rewrite the corresponding
equations in the spherical limit. NN correlations modify the high-$q$ tails 
of the form factors ($q$ beyond 2 fm$^{-1}$), improving the agreement with experiment.
On the other hand, as we shall see in the next section, the deformed formalism is
needed to describe properly the measured form factors, as it was also necessary to
account for the values of the experimental magnetic moments in deformed nuclei.

\subsection{Deformed nuclei}

In the next figures, Figs. \ref{fig_25mg_ff_sph_def}-\ref{fig_115in_ff_sph_def}, we 
compare in the top panels calculations from the deformed and spherical formalisms in 
the case of deformed nuclei, that according to Table \ref{table1} are $^{25}$Mg, 
$^{29}$Si, $^{59}$Co, and $^{115}$In. In this case, to calculate the spherical limit a 
HF+BCS calculation constrained to zero deformation has been performed for each nucleus.
In the middle panels we show the total magnetic form factor from the deformed model
decomposed into multipolarities with single-particle contributions only. The bottom 
panels show the results including rotational collective contributions from the core  
${\cal F}^{M\lambda}_R$, calculated in the cranking model. Explicit expressions for these 
contributions can be seen in Ref. \cite{elvira1}. We studied in the past \cite{ps,berdi} 
the effect of these collective rotational contributions from different microscopic and 
macroscopic models and concluded that they are, in general, small compared to 
single-particle contributions. They are only expected to show up in the M1 multipoles 
at low $q$ and do not differ much from one rotational model or another. Thus, we opted 
here for calculations from the cranking model that produce better moments of inertia.

Figure \ref{fig_25mg_ff_sph_def} shows the results for $^{25}$Mg ($I^{\pi}=5/2^+$) with 
data taken from \cite{donnelly,york,euteneuer}. In the top panel we show the deformed 
calculations compared with the spherical limit, as well as with the standard spherical 
calculation with CDFM correlations. We observe in the deformed case a clear improvement 
of the agreement with experiment. The first peak is due to M1, whereas the second is 
due to M5, with a negligible contribution from M3. We can see that the contribution of 
the core rotational currents appears mainly in the first peak, due to the effect on M1 
at low $q$. This contribution improves the agreement in the first peak. The high $q$ 
behavior is determined by the M5 multipole, which is practically unaffected by  
${\cal F}^{M5}_R$.

In the next figure, Fig. \ref{fig_29si_ff_sph_def}, we show the results for $^{29}$Si 
($I^{\pi}=1/2^+$). In this case only the M1 multipole contributes and one panel is enough 
to show the various contributions. We get a much better agreement with data 
\cite{donnelly} in the deformed case. We should note that the improvement is due 
directly to the structure of the deformed orbital in comparison to the spherical, 
because in this case the M1 multipole is not reduced by the geometrical factor 
$\eta_j^{\lambda}$. There is practically no contribution from the core rotational current, 
even though in this case we have an extra contribution from the core proportional to 
the decoupling parameter $a=1.166$ (see Eq. (\ref{fm2k})). 

Figure \ref{fig_59co_ff_sph_def} displays the results for  the nucleus $^{59}$Co 
($I^{\pi}=7/2^-$). The data from  \cite{donnelly,devries70,platchkov82} are better 
reproduced by the spherical case at low and high $q$ values, but the deformed 
calculations reproduce better the whole structure of the first peak including the 
fall of the curve. The filling of the form factors in the range $1<q<2$ fm$^{-1}$ 
produced by the spherical calculation is not observed experimentally, which seems to 
favor the deformed picture. This could be an indication that we are dealing in this 
case with small deformations, see Table \ref{table1}. We can see again a small
contribution from the core rotational current in the first peak that improves 
slightly the agreement with experiment.

In the case of  $^{115}$In ($I^{\pi}=9/2^+$), Fig. \ref{fig_115in_ff_sph_def}, the
data from \cite{donnelly} seem to show three peaks, which are better described in 
the deformed picture. The spherical description smears too much the profile of the 
curve. In this nucleus, collective effects do not play a significant role.

It is worth noting that in the cases of $^{59}$Co and $^{115}$In, the form factors at
low $q$ that determine the magnetic moment of the nucleus are quite similar in the
cases of spherical and deformed calculations, with somewhat larger values in the
spherical case. This is correlated to the values of the magnetic moments quoted in
Table I. We can see that the magnetic moments from spherical and deformed calculations
are close to each other with slightly larger values in the spherical case. On the
other hand, in the cases of  $^{25}$Mg and $^{29}$Si, the behavior of M1 at low $q$
is very different, as well as the corresponding magnetic moments in Table I.

In summary, we have found that the geometrical factors in the deformed formalism
reduce the multipoles with respect to the spherical ones and help to improve the 
agreement with experiment in all the deformed cases studied.
These reduction factors appear naturally in the deformed formalism, whereas they 
have to be introduced {\it ad hoc} to fit the data in other approaches \cite{dong,wang15}.
Collective effects manifest mainly at low-$q$ values through M1 rotational multipoles.
Although they are rather small in these calculations, they improve the 
agreement with experiment.

\section{Conclusions}

In this paper we have calculated magnetic form factors in elastic electron scattering 
from odd-$A$ nuclei within PWBA and within a deformed formalism, using  wave functions 
from self-consistent HF+BCS calculations with Skyrme forces.

We have recovered the spherical limit of these calculations and have compared the
results with those obtained from spherical codes, finding a perfect agreement. We have 
shown that spherical nuclei are well described in this limit. NN correlations
are included with natural orbits within the CDFM and are found to shift the tails of 
the form factors to higher momentum transfer, improving the agreement with experiment.

Then we proceed to calculate deformed nuclei and compare both spherical and deformed
calculations with experiment. For these nuclei ($^{25}$Mg, $^{29}$Si, $^{59}$Co, and 
$^{115}$In), the deformed picture represents clearly an improvement of the agreement 
with the data, demonstrating the need for deformation degrees of freedom to describe 
these nuclei. The role of the collective rotation current of the core has been studied, 
showing that it manifests itself mainly in the M1 multipole, changing the profile of 
the form factor at low momentum transfer and improving in general the agreement with 
the measurements.

It is found that in odd-$A$ deformed nuclei, the main contribution to the magnetic 
form factor comes from the odd particle. This result is in contrast to the case of 
charge form factors, where all the nucleons contribute significantly. The collective 
effects in the deformed formulation enter in three different ways: i) the deformation 
that modifies the single-particle wave functions with respect to the wave functions 
in the spherical case; ii) the collective current contributions through the rotational 
multipoles  ${\cal F}^{M\lambda}_R$; and iii) the reduction of the single-particle 
multipole contributions with respect to the spherical ones due to the strong 
particle-rotor coupling that manifests itself in the so-called geometrical factors.
The effect of the rotational multipoles is more important in the low-$q$ region 
($q<1$ fm$^{-1}$), where they interfere with the single-particle contributions.
This collective contribution from the deformed core must be included for a quantitative 
comparison with experiment.

We have shown that we can deal with spherical and deformed isotopes in a unified way 
using the same numerical methods and codes.  It is also worth noting that, at variance 
with the approach followed in other works \cite{dong,wang15}, in this paper there is 
no fit of the coefficients weighting the various multipoles contributing to the total 
magnetic form factors. In the present formalism, the weights of the multipoles are 
directly given by the geometrical factors relating the intrinsic with the transition 
multipoles and therefore we do not introduce any adjustable parameter. 

Once the capability of the model has been tested against data on magnetic form factors
on spherical and deformed stable nuclei, we have a trustable formalism to explore
the predictions on unstable nuclei.

\begin{acknowledgments}
This work was supported by Ministerio de Ciencia, Innovaci\'on y Universidades (Spain) 
under Contract Nos. FIS2014-51971-P and PGC2018-093636-B-I00.  Three of the authors 
(M.K.G., A.N.A., and D.N.K.) are grateful for support of the Bulgarian Science Fund 
under Contract No. DFNI-T02/19.

\end{acknowledgments}

\end{document}